# Adaptive System Identification using Markov Chain Monte Carlo

**Muhammad Ali Raza Anjum**
Department of Engineering Sciences,
Army Public College of Management and Sciences, Rawalpindi, PAKISTAN
e-mail: ali.raza.anjum@apcoms.edu.pk

*Abstract*

One of the major problems in adaptive filtering is the problem of system identification. It has been studied extensively due to its immense practical importance in a variety of fields. The underlying goal is to identify the impulse response of an unknown system. This is accomplished by placing a known system in parallel and feeding both systems with the same input. Due to initial disparity in their impulse responses, an error is generated between their outputs. This error is set to tune the impulse response of known system in a way that every change in impulse response reduces the magnitude of prospective error. This process is repeated until the error becomes negligible and the responses of both systems match. To specifically minimize the error, numerous adaptive algorithms are available. They are noteworthy either for their low computational complexity or high convergence speed. Recently, a method, known as Markov Chain Monte Carlo (MCMC), has gained much attention due to its remarkably low computational complexity. But despite this colossal advantage, properties of MCMC method have not been investigated for adaptive system identification problem. This article bridges this gap by providing a complete treatment of MCMC method in the aforementioned context.

## 1. Introduction

System identification is one of the most important problems in adaptive filtering. Over the past many years, it has been studied extensively in many areas of science and engineering, both in theoretical and applied contexts. For example, volumes of results in this area have been reported for echo cancellation in acoustics [1], for channel estimation in communications [2], for earth-sounding in geophysics [3], for robotics and control [4], etc. More and more sophisticated techniques are perpetually being reported in this area [5-11]. Therefore, its importance can be hardly over-emphasized.

In system identification problem, the impulse response of an unknown system is identified by placing a discrete-time Finite Impulse Response (FIR) system in parallel with the unknown system and feeding both systems with same input. The difference between the outputs of two systems is regarded as error and the response of the FIR system is adjusted minimize the error. This process is repeated until the error in the outputs of both systems becomes negligible. When this happens, the impulse responses of both systems match. This, in very general terms, is the principle behind the system identification problem [12].

The principal task associated with system identification problem is minimization of the indicated error. One way to do it is to reduce average power of the error by forming its cost function. This is regarded as the principle of Minimum Mean Square Error (MMSE). Average power of the error, also known as mean squared error, is taken as the cost of selecting a particular impulse response for the FIR system. The response is selected in such a way that this cost is minimized over a set of all possible selections provided further reduction in cost is not possible. Hence, the system identification problem is transformed into an optimization problem [13].

There are many iterative algorithms that can search the cost function for such an optimal impulse response. Least Mean Squares (LMS) algorithm, Normalized LMS (NLMS), and Recursive Least Squares (RLS) algorithm are very famous in this regard. The distinguishing points for any such algorithm are its computational complexity and speed of convergence. Both are generally considered as tradeoffs. For example, LMS algorithm has very low computational complexity but is very slow to converge. RLS algorithm, on the contrary, has rapid convergence properties but is computationally intensive [12].

Recently, a very interesting method, known by the name of Markov Chain Monte Carlo (MCMC) method, has emerged in the computing literature for the solution linear algebraic equations [14]. Though the history of the MCMC method can be traced as far back as 1950s [15], it has gained much attention over recent years due to the availability of powerful computing devices. This method employs random number generators for the solution of practical problems where a closed form or analytical solution is not possible. However, this method has also been studied lately for conventional problems due to its very low computational complexity [16-19].

This feature also makes it an excellent choice for the study of system identification problem in adaptive filtering. So far as can be ascertained, MCMC method and its properties have not been investigated in this context. This article aims to fill this gap and, hence, provides a complete treatment of MCMC method for adaptive system identification problem.

**2. Nomenclature**

Consider the system identification problem depicted in Figure 1.

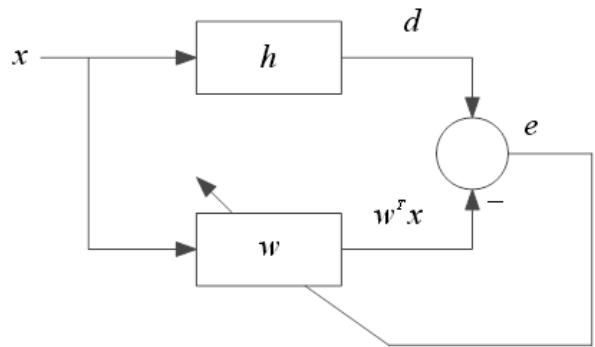

Figure 1. Nomenclature for system identification problem

Let there be a discrete-time Linear Time Invariant (LTI) system with a finite impulse response $h$ of length $N$.

$$\boldsymbol{h} = [h_0 \quad h_1 \quad ... \quad h_{N-1}]^T \tag{1}$$

$h$ is unknown. To identify it, a Finite Impulse Response (FIR) filter $w$ of equal length is placed parallel to the unknown system.

$$\boldsymbol{w} = [w_0 \quad w_1 \quad ... \quad w_{N-1}]^T \tag{2}$$

A Wide Sense Stationary (WSS) random sequence $x$ is applied to the input of both systems.

$$\boldsymbol{x} = [x_0 \quad x_1 \quad ... \quad x_{N-1}]^T \tag{3}$$

$x$ has following properties.

$$E\{x_i\} = 0 \tag{4}$$

And,

$$E\{x_i^2\} = \sigma_x^2 \tag{5}$$

$\sigma_x^2$ is the power of the input sequence $x$. It is generally normalized to unity.

$$\sigma_x^2 = 1 \tag{6}$$

For the given input $x$, the unknown system generates an output $h^T x$ and the FIR filter generates an output $w^T x$. Former is termed as the desired input $d$.

$$d = h^T x \tag{7}$$

Difference between the desired output and the filter output is termed as error $e$.

$$e = d - w^T x \tag{8}$$

### 3. Derivation of Wiener-Hopf Equation

Derivation of Wiener-Hopf Equation can be a two step process. In first step, a cost function of the error is formed by squaring it and then taking its expected value.

$$E\{e^2\} = d^2 - 2w^T E\{xd\} + w^T E\{xx^T\}w \tag{9}$$

Or,

$$\xi = d^2 - 2w^T b + w^T R w \tag{10}$$

$\xi$ is the cost function. It represents the Mean Square Error (MSE). $p$ represents the cross-correlation vector.

$$b = [b_0 \quad b_1 \quad ... \quad b_{N-1}]^T \tag{11}$$

$R$ is the auto-correlation matrix.

$$R = \begin{bmatrix} r_{00} & r_{01} & ... & r_{0(N-1)} \\ r_{10} & r_{11} & ... & r_{1(N-1)} \\ \vdots & \vdots & \ddots & \vdots \\ r_{(N-1)0} & r_{(N-1)1} & ... & r_{(N-1)(N-1)} \end{bmatrix} \tag{12}$$

Since $x$ is WSS, $R$ is a symmetric Toeplitz matrix [13].

$$r_{ij} = r_{i-j} \tag{13}$$

And,

$$r_{i-j} = r_{j-i} \tag{14}$$

So,

$$R = \begin{bmatrix} r_0 & r_1 & ... & r_{(N-1)} \\ r_1 & r_0 & ... & r_{(N-2)} \\ \vdots & \vdots & \ddots & \vdots \\ r_{(N-1)} & r_{(N-2)} & ... & r_0 \end{bmatrix} \tag{15}$$

In second step, the cost function in Eq. (10) is minimized by taking its gradient with respect to $w$ and then setting the gradient to zero.

$$0 = -2b + 2Rw \tag{16}$$

Or,

$$Rw = b \tag{17}$$

Eq. (17) represents the famous Weiner-Hopf Equation. $w$ is known as the Minimum MSE (MMSE) Wiener Filter.

## 4. Markov Chain Monte Carlo Solution of the Wiener-Hopf Equation

In Eq. (16), $R$ matrix can be split as,

$$R = I - F \tag{18}$$

So that,

$$(I - F)w = b \tag{19}$$

Or,

$$w = (I - F)^{-1}b \tag{20}$$

But,

$$(I - F)^{-1} = I + F + F^2 + \cdots \tag{21}$$

Substituting Eq. (21) in Eq. (20),

$$w = (I + F + F^2 + \cdots)b \tag{22}$$

Expanding Eq. (22),

$$w = b + Fb + F^2b + \cdots \tag{23}$$

Further expanding the $F$ matrix in Eq. (22),

$$F = \begin{bmatrix} f_{00} & f_{01} & \cdots & f_{0(N-1)} \\ f_{10} & f_{11} & \cdots & f_{1(N-1)} \\ \vdots & \vdots & \ddots & \vdots \\ f_{(N-1)0} & f_{(N-1)1} & \cdots & f_{(N-1)(N-1)} \end{bmatrix} \tag{24}$$

Using Eqs. (24), (2) and (11) to expand Eq. (22),

$$\begin{bmatrix} w_0 \\ w_1 \\ \vdots \\ w_{N-1} \end{bmatrix} = \begin{bmatrix} b_0 \\ b_1 \\ \vdots \\ b_{N-1} \end{bmatrix} + \begin{bmatrix} f_{00} & f_{01} & \cdots & f_{0(N-1)} \\ f_{10} & f_{11} & \cdots & f_{1(N-1)} \\ \vdots & \vdots & \ddots & \vdots \\ f_{(N-1)0} & f_{(N-1)1} & \cdots & f_{(N-1)(N-1)} \end{bmatrix} \begin{bmatrix} b_0 \\ b_1 \\ \vdots \\ b_{N-1} \end{bmatrix} + \cdots \tag{25}$$

Picking the $i$-th unknown in Eq. (24),

$$w_i = b_i + \sum_{k=0}^{N-1} f_{ik} b_k + \sum_{k=0}^{N-1}\sum_{j=0}^{N-1} f_{ij} f_{jk} b_k + \cdots \tag{26}$$

Splitting $f$'s in Eq. (26),

$$f_{ij} = p_{ij} v_{ij} \tag{27}$$

Where $v_{ij} \in \mathbb{R}$ and,

$$0 \le p_{ij} \le 1 \tag{28}$$

$p_{ij}$'s represent probabilities. Substituting Eq. (27) in Eq. (26),

$$w_i = b_i + \sum_{k=0}^{N-1} p_{ik} v_{ik} b_k + \sum_{k=0}^{N-1}\sum_{j=0}^{N-1} p_{ij} p_{jk} v_{ij} v_{jk} b_k + \cdots \tag{29}$$

Or,

$$w_i = \lim_{m \to \infty} s_{im} \tag{30}$$

$s_{im}$ represents the sum of first $m$ terms. $s_{i1}$ term comes from the $i$-th row of $I$, $s_{i2}$ comes from the $i$-th rows of $I$ and $F$, $s_{i3}$ comes from the $i$-th rows of $I, F, F^2$ and so on.

### 4.1. Analysis of the Infinite Sum

Decomposing the $s_{i1}$ term,

$$s_{i1} = b_i \tag{31}$$

$s_{i1}$ is equal to the $i$-th value of the right-hand side in Eq. (17). We remember it as the starting position. Decomposing the $s_{i2}$ term,

$$s_{i2} = b_i + p_{i0} v_{i0} b_0 + p_{i1} v_{i1} b_1 + \cdots + p_{i(N-1)} v_{i(N-1)} b_{N-1} \tag{32}$$

Let,

$$p_{ij} = \frac{1}{N} \tag{33}$$

With $j = 0, 1, N - 1$. Substituting Eq. (33) in Eq. (32),

$$s_{i2} = b_i + \frac{1}{N} v_{i0} b_0 + \frac{1}{N} v_{i1} b_1 + \cdots + \frac{1}{N} v_{i(N-1)} b_{N-1} \tag{34}$$

Re-arranging Eq. (34),

$$s_{i2} = (Nb_i + v_{i0}b_0 + v_{i1}b_1 + \cdots + v_{i(N-1)}b_{N-1})/N \tag{35}$$

$s_{i2}$ term appears to be an average. To analyze it further, we move the $N$ in the denominator of Eq. (35) to left.

$$Ns_{i2} = Nb_i + v_{i0}b_0 + v_{i1}b_1 + \cdots + v_{i(N-1)}b_{N-1} \tag{36}$$

Decomposing Eq. (36) into $N$ equations,

$$\begin{aligned} s_{i2} &= b_i + v_{i0}b_0 \\ s_{i2} &= b_i + v_{i1}b_1 \\ &\vdots \\ s_{i2} &= b_i + v_{i(N-1)}b_{N-1} \end{aligned} \tag{37}$$

These $N$ equations represent $N$ one-step random walks taken by the $i$-th particle. In first equation for example, the particle starts from state-$i$ (characterized by $b_i$), moves to state-0 (characterized by $b_o$), and then comes to a halt. There are $N$ one-step terminating points $v_{ik}b_k$. Hence, $s_{i1}$ is an average of $N$ one-step random walks of the $i$-th particle to $N$ one-step terminating points $v_{ik}b_k$ when it starts from state-$i$. In the light of this discussion, we analyze the term $s_{i3}$.

$$s_{i3} = s_{i2} + \sum_{k=0}^{N-1}\sum_{j=0}^{N-1} p_{ij}p_{jk}v_{ij}v_{jk}b_k \tag{38}$$

Let,

$$p_{ij} = p_{jk} = \frac{1}{N} \tag{39}$$

For $\forall\, j, k$. Substituting Eq. (39) in Eq. (38),

$$s_{i3} = s_{i2} + \sum_{k=0}^{N-1}\sum_{j=0}^{N-1} \frac{1}{N^2} v_{ij}v_{jk}b_k \tag{40}$$

$s_{i3}$ contains an additional double summation term when compared to $s_{i2}$. This extra term represents two-step random walks of the $i$-th particle to two-step terminating points $v_{ij}v_{jk}b_k$ before it comes to a halt. There will be $N^2$ such random walks. Hence, $s_{i3}$ is an average of $N^2$ two-step random walks of the $i$-th particle to $N^2$ two-step terminating points $v_{ij}v_{jk}b_k$ when it starts from state-$i$. We can show in a similar way that $s_{i4}$ is an average of $N^3$ three-step random walks to $N^3$ three-step terminating points $v_{il}v_{lj}v_{jk}b_k$. Continuing in a similar manner, $s_{im}$ is an average of $N^{m-1}$ $(m-1)$-step random walks to $N^{m-1}$ terminating points.

**4.2. Random Walks with Arbitrary Probabilities**

In Eq. (32), we assumed,

$$p_{ik} = \frac{1}{N} \tag{Repeat}$$

With $k = 0, 1, …, N − 1$. This means when a particle is in state-$i$, it is equally likely to transit to any of the $N$ one-step terminating points with a probability $1/N$. This implies that $N$ one-step terminating points will be covered in at least $N$ random walks, $N^2$ two-step terminating points will be covered in at least $N^2$ random walks, $N^3$ three-step terminating points will be covered in at least $N^3$ random walks, and so on. But if the probabilities are not equally likely, this will not be the case. We will have to define another variable $M^{(i)}$ to indicate the number of minimum random walks that will at least be required to cover $i$-step terminating points. In one-step terminating points, $M^{(1)}$ will be inversely proportional to the smallest transition probability among all the one-step transition probabilities from state-$i$ to all one-step terminating points. This probability is then rounded off upwards to the nearest possible integer.

$$M^{(1)} = \left\lceil \frac{1}{\min\limits_{\forall k} p_{ik}} \right\rceil \tag{41}$$

With $k = 0, 1, …, N − 1$. Eq. (41) means that $k$-th one-step terminating point with the smallest transition probability $p_{ik}$ will be covered in at least $1/p_{ik}$ random walks. Similarly, $M^{(2)}$ will be equal to the smallest transition probability, rounded upwards to the nearest possible integer, among all the two-step transition probabilities from state-$i$ to all two-step terminating points.

$$M^{(2)} = \left\lceil \frac{1}{\min\limits_{\forall j,k} p_{ij} p_{jk}} \right\rceil \tag{42}$$

With $j, k = 0, 1, …, N − 1$. Similarly, $M^{(i)}$ can be written as,

$$M^{(i)} = \left\lceil \frac{1}{\min\limits_{\forall j,h,…,l,k} p_{ij} p_{jh} … p_{lq} p_{qk}} \right\rceil \tag{43}$$

With $j, h, …, l, q, k = 0, 1, …, N − 1$.

### 5. Analysis of Convergence

Our entire analysis started from Eq. (21) which represents an infinite matrix series.

$$I + F + F^2 + \cdots = \sum_{j=0}^{\infty} F^j = \lim_{k \to \infty} \sum_{j=0}^{k} F^j = \lim_{k \to \infty} U_k \tag{44}$$

Where,

$$U_k = \sum_{j=0}^{k} F^j \tag{45}$$

Let there be a solution $U$ to which $U_k$ will converge in limit.

$$U = \lim_{k \to \infty} U_k \tag{46}$$

But,

$$U_k = \sum_{j=0}^{k-1} F^j + F^k = U_{k-1} + F^k \tag{47}$$

Or,

$$F^k = U_k - U_{k-1} \tag{48}$$

Applying the limit,

$$\lim_{k \to \infty} F^k = \lim_{k \to \infty} U_k - \lim_{k \to \infty} U_{k-1} = U - U = 0 \tag{49}$$

Or,

$$\lim_{k \to \infty} F^k = 0 \tag{50}$$

Eq. (50) implies that,

$$\|F\| < 1 \tag{51}$$

Therefore, a canonical norm of $F$ matrix should be less than unity for the convergence of the MCMC method to a unique solution. Presence of a unique solution will in turn guarantee the stability of the method by Lax's Theorem [20].

### 6. Limitations of the Method

Eq. (51) imposes a restriction on $F$ matrix. One way to analyze this restriction is to look at the eigenvalues of $F$ matrix. They should lie inside the following interval to meet this restriction.

$$-1 < \lambda_i^F < 1 \tag{52}$$

Complex eigenvalues can equally meet the condition described in Eq. (52). But we will focus our attention on real eigenvalues because $F$ has real eigenvalues. It is related to $R$ matrix by Eq. (17). As $R$ is a symmetric matrix [13], it will have real eigenvalues by fundamental theorem of linear algebra [21]. Hence, $F$ will also be a symmetric matrix with real eigenvalues. By the relation described by Eq. (17), condition on eigenvalues of $F$ matrix translates directly to the following condition on eigenvalues of $R$ matrix.

$$0 < \lambda_i^R < 2 \tag{53}$$

Eq. (53) means that all eigenvalues of $R$ should be real, positive, and inside the interval [0,2]. For $R$ matrix, first two conditions hold automatically. $R$ is a symmetric matrix so its eigenvalues are real. $R$ is positive definite so all its eigenvalues are positive [13]. Finally, to prove that $R$ meets the third constraint as well, we employ Gershgorin's theorem [22]. This theorem states that $i$-th eigenvalue of a matrix $A = \{a_{ij}\}$ lies inside a circle,

$$C_i\left(a_{ii}, \sum_{i \neq j} |a_{ij}|\right) \qquad (54)$$

such that $a_{ii}$ is the centre and $\sum_{i \neq j}|a_{ij}|$ is the radius of corresponding circle. There will be $N$ such circles for an $N \times N$ matrix. Since $R$ is a finite symmetric Toeplitz matrix, all the Gershgorin's circles will overlap such that,

$$C\left(r_0, \sum_{i=1}^{N-1} |r_i|\right) \qquad (55)$$

$r_n$ represents the autocorrelation sequence of the input random process $x$. $r_o$ is its value at zero delay. This is equivalent to the energy of the random process $x$ which can be normalized to unity. So the Gershgorin's circles for $R$ will be centered at unity. As to the remaining values of $r_n$, its monotonically decreasing nature will ensure that the remaining values lie inside a certain region [23]. For convergence, this region must be bounded by Eq. (55). For example, if the samples of $x$ are independent and identically distributed (IID) with zero mean and unity variance, then,

$$E\{r_i r_j\} = \begin{cases} 1 & i = j \\ 0 & i \neq j \end{cases} \qquad (56)$$

There will be $N$ overlapping Gershgorin's circles $C(1,0)$. These will essentially be $N$ points centered at 1. So, there will be $N$ eigenvalues equal to 1. This is true because by Eq. (17), $R$ will be equivalent to an identity matrix which has all eigenvalues equal to 1. In this case, the starting positions of the $N$-particles, which are represented by $b$, will be the solution to the system identification problem.

$$w = b \qquad (57)$$

This once again is true because when $R$ is an identity matrix, $w = b$.

### 7. Analysis of Error

We begin with the infinite series described by Eq. (30).

$$w_i = \lim_{m \to \infty} s_{im} \qquad \text{(Repeat)}$$

For $m = 1$, $w_i$ will be equal to the first term in the infinite series.

$$w_i = s_{i1} = b_i \qquad \text{(Repeat)}$$

This will be the starting value of the particle. So error in the estimate at the beginning of the random walk will be equal to,

$$e_0 = w_i - s_{i1} = \lim_{m \to \infty} s_{im} - s_{i1} \tag{58}$$

0 indicates that no random walks have been initiated yet. For $m = 2$, $w_i$ will be equal to the sum of first two term in the infinite series.

$$w_i = s_{i2} = b_i + \sum_{k=0}^{N-1} p_{ik} v_{ik} b_k \tag{Repeat}$$

This represents all one-step terminating points. A minimum of $M^{(1)}$ random walks will be required to cover all these points. So the error after $M^{(1)}$ random walks will be,

$$e_{M^{(1)}} \geq \lim_{m \to \infty} s_{im} - s_{i2} \tag{59}$$

For $m = 3$,

$$w_i = s_{i3} = b_i + \sum_{k=0}^{N-1} p_{ik} v_{ik} b_k + \sum_{k=0}^{N-1} \sum_{j=0}^{N-1} p_{ij} p_{jk} v_{ij} v_{jk} b_k \tag{Repeat}$$

This represents all two-step terminating points. A minimum of $M^{(2)}$ random walks will be required to cover all these points. So the error after $M^{(2)}$ random walks will be,

$$e_{M^{(2)}} \geq \lim_{m \to \infty} s_{im} - s_{i3} \tag{60}$$

Similarly, the error after $M^{(j)}$ random walks will be,

$$e_{M^{(j)}} \geq \lim_{m \to \infty} s_{im} - s_{ij} \tag{61}$$

$e_{M^{(j)}}$ represents the lower bound on error. In general, the error in the estimate after $M^{(j)}$ random walks will be greater than $e_{M^{(j)}}$. This is due to the law of large numbers which dictates that $M^{(j)} \to \infty$ for the equality to hold in Eq. (61). Also,

$$e_0 > e_{M^{(1)}} > e_{M^{(2)}} \ldots > e_{M^{(j)}} \tag{62}$$

When $j \to \infty$,

$$e_{M^{(\infty)}} = \lim_{m \to \infty} s_{im} - \lim_{j \to \infty} s_{ij} = 0 \tag{63}$$

Hence, infinite random walks will be required to take the error in the estimate equal to zero.

## 8. Construction of Algorithm

In this section, we focus our attention on the construction of an algorithm to compute an iterative MCMC solution to Wiener-Hopf Equation.

### 8.1. The Matter of Absorbing State

From an algorithmic viewpoint, there is one essential point to consider. For that purpose, let us analyze the second random walk in Eq. (37). The particle starts from state-$i$, moves to state-1 in next step and then the random walk suddenly comes to an end. So in addition to the existing $N$ states, there must be another state, a no-return state. We term this extra state as the absorbing state. A particle cannot return from absorbing state.

$$\sum_{k=1}^{N-1} p_{Ni} = 0 \qquad (64)$$

And once the particle encounters the absorbing state, it stays there.

$$p_{NN} = 1 \qquad (65)$$

Hence for a single unknown, there are a total of $(N + 1)$ states. Associated with these states are $(N + 1)$ state transition probabilities.

### 8.2. Formation of the State Transition Matrix

Each unknown has $N + 1$ state transition probabilities associated with it. As there are $N$ unknowns in the system, we can construct an $N \times (N + 1)$ state transition matrix $\boldsymbol{P}$.

$$\boldsymbol{P} = \begin{bmatrix} p_{00} & p_{01} & \cdots & p_{0(N-1)} & p_{0N} \\ p_{10} & p_{11} & p_{12} & p_{1(N-1)} & p_{1N} \\ \vdots & \vdots & \vdots & \vdots & \vdots \\ p_{(N-1)0} & p_{(N-1)1} & \cdots & p_{(N-1)(N-1)} & p_{(N-1)N} \end{bmatrix} \qquad (66)$$

Probabilities associated with absorbing state are called absorption probabilities, i.e., $p_{iN} \; \forall \; i = 0,1,\ldots,N-1$. We add an extra row to $\boldsymbol{P}$ in order to meet the conditions laid down in Eqs. (64) and (65) for the absorbing state.

$$\boldsymbol{P} = \begin{bmatrix} p_{00} & p_{01} & \cdots & p_{0(N-1)} & p_{0N} \\ p_{10} & p_{11} & p_{12} & p_{1(N-1)} & p_{1N} \\ \vdots & \vdots & \vdots & \vdots & \vdots \\ p_{(N-1)0} & p_{(N-1)1} & \cdots & p_{(N-1)(N-1)} & p_{(N-1)N} \\ 0 & 0 & 0 & 0 & 1 \end{bmatrix} \qquad (67)$$

The dimensions of the $\boldsymbol{P}$ matrix now change to $(N + 1) \times (N + 1)$. The absorption probabilities will not be available directly. In order to obtain them, we should ensure during splitting process in Eq. (27) that,

$$\sum_{k=0}^{N-1} p_{ik} < 1 \qquad (68)$$

$\forall \; i = 0,1,\ldots,N-1$. By fundamental theorem of probability [24],

$$\sum_{k=0}^{N} p_{ik} = 1 \tag{69}$$

From Eq. (69), absorption probabilities can be computed in the following manner.

$$p_{iN} = 1 - \sum_{k=0}^{N-1} p_{ik} \tag{70}$$

$\forall\ i = 0,1,\dots,N-1$.

### 8.3. Formation of State Transition Rules
Once the state transition matrix is complete, state transition rules for the $i$-th unknown $w_i$ can be defined according to Table 1.

**Table 1.** State transition rules for $i$-th unknown

| Probability value | State |
| --- | --- |
| $p < p_{i0}$ | State-0 |
| $p_{i0} \leq p < (p_{i0} + p_{i1})$ | State-1 |
| ⋮ | ⋮ |
| $\sum_{j=0}^{N-1} p_{ij} \leq p$ | State-N |

$p$ is a random number drawn from a uniform distribution over the interval $[0,1]$.

### 8.4. Algorithm
Now we present an iterative algorithm to compute an arbitrary unknown according to MCMC method.

1. Split $R$ into $(I - F)$ according to Eq. (18).
2. Split $f$'s according to Eq. (27).
3. Form the state transition matrix $P$ according to Eqs. (66)-(70).
4. Select the unknown to be computed.
5. Define the state transition rules for that unknown according to Table 1.
6. Start random walks by drawing random numbers from a uniform probability distribution over $[0,1]$.
7. Follow state transition rules to assign values to the unknown according to Table 1.
8. Return to step six and keep iterating until a given number of iterations is met.
9. Compute the average of all the random walks.

The average will provide the estimated value of the selected unknown. Similar procedure can be repeated to compute the estimates of remaining unknowns.

## 9. Computational Complexity of the Algorithm

In this section, we analyze the computational complexity of the algorithm. One way to measure the computational complexity of an iterative algorithm is to compute the number of multiplications required by the algorithm in a single iteration [2]. The MCMC algorithm requires one multiplication per iteration for the computation of a single unknown, .i.e., multiplication of $b$'s with $v$'s as indicated by Eqs. (27) and (29). Since the algorithm is iterative, the multiplication of $b$'s with $v$'s in each step will be carried on to the next step. The next step will again require a single similar multiplication and so on. To compute all the $N$ unknowns in the unknown vector $w$, $N$ multiplications per iteration will be required. Hence, the order of complexity of the algorithm is proportional to $N$. Table 1 compares the computational complexity of the MCMC algorithm with state of the art algorithms available in literature for iterative solution of Wiener-Hopf equation [12]. As can be observed from the table, MCMC algorithm has the lowest computation complexity among all its competitors.

Table 2. Comparison of the computational complexity of the Monte Carlo algorithm

| Algorithm | Complexity |
|---|---|
| MCMC | $N$ |
| LMS | $2N + 1$ |
| NLMS | $3M + 2$ |
| Kaczmarz | $3M + 2$ |
| RLS | $N^2$ |

## 10. Simulation Results and Conclusion

Now we present the simulation results of the MCMC algorithm and compare them with the results obtained for LMS, NLMS, and RLS algorithms. A two tap system identification problem is selected, i.e., $N = 2$. There will be two transient states and one absorbing state for this problem. This will lead to a $3 \times 3$ probability transition matrix with two tables of transition rules defined for the two unknowns according to Table1. The MCMC algorithm is run for a certain number iterations to generate the simulation results in terms of second norm of the error in the estimate. Results are summarized in Table 3. According to Table 3, the MCMC algorithm yields much lower error norm than LMS algorithm but much higher norm as compared to NLMS and RLS algorithms. This indicates that the MCMC algorithm is relatively slow to converge. This has been expected from error analysis in section 7. The slow convergence of the algorithm can be attributed to sluggishness of the averaging process inherent to the MCMC method. However, this tardiness in its convergence is offset by its nominal computational complexity. Hence, the MCMC method can be recommended for practical situations where a rough estimate of the unknown system is desired due to limited power or computing resources.

Table 3. Comparison of simulation results

| Iterations | MCMC | LMS $\mu = 0.01$ | NLMS | RLS |
|---|---|---|---|---|
| | $\|e\|$ | $\|e\|$ | $\|e\|$ | $\|e\|$ |
| 02 | 0.0839 | 1.3424 | 0.5000 | 0.5000 |
| 04 | 0.0768 | 1.3402 | 0.2023 | 0.0549 |
| 08 | 0.0596 | 1.2761 | 0.0607 | 9.4485e-004 |
| 16 | 0.0613 | 1.2461 | 3.3428e-004 | 5.6911e-009 |
| 32 | 0.0558 | 1.1123 | 9.6127e-013 | 1.6012e-015 |
| 64 | 0.0576 | 0.9160 | 0 | 0 |